# Spectral Fingerprints of Street-Network Morphology: A Size-Adjusted Graph-Laplacian Descriptor of Urban Fabric

**Piotr C. Kamiński** *[Magdalena Abakanowicz University of the Arts: Poznań, PL; ORCID: 0009-0008-6573-2057] Correspondence: [piotr.kaminski@uap.edu.pl]*

---

## Abstract

Decades of space-syntax research have established that the topology of the street network conditions movement, co-presence and the distribution of urban activity. The standard analytical vocabulary for this - integration, choice and connectivity - summarises each street's position as a scalar centrality. Two streets with identical centrality can however sit in radically different morphological fabric. This paper introduces a compact, comparable and machine-learning-ready encoding of that local fabric: the *spectral fingerprint*, a fixed-dimensional kernel-density representation of the graph-Laplacian eigenvalue distribution of each node's k-hop ego subgraph, computed on the COINS dual graph of the street network. From the fingerprint we derive two interpretable scalar readouts: the *Mesh Index* (MI), a normalised spectral entropy, and the *Connectivity Resilience Index* (CRI), the algebraic connectivity (Fiedler value). Both are corrected for an ego-subgraph-size confound that otherwise dominates raw spectral statistics. Applied to the full street network of the city of Poznań, Poland (1,908 continuity-based strokes), the descriptor is shown to be robust to its encoding hyperparameters (spectral resolution and kernel bandwidth; Spearman $\rho \geq 0.98$) while remaining genuinely scale-dependent in its neighbourhood radius. The size adjustment removes a strong confound, leaving the Mesh Index near-orthogonal to integration ($r = 0.06$) and thus carrying information that classical centrality does not. The fingerprint separates morphological tissue types without supervision, and the Mesh Index is associated with the functional diversity of street-adjacent activity at the neighbourhood scale ($r \approx 0.19$, on open OpenStreetMap data). We delineate the method's scope honestly: it characterises what kind of activity a street's position affords, not the price that activity commands. The spectral fingerprint offers an information-theoretic morphological descriptor that complements space syntax and is directly consumable by modern graph-learning pipelines.

**Keywords:** spectral graph theory; graph Laplacian; spectral entropy; space syntax; urban morphology; street networks

---

## 1. Introduction

The street network is the most persistent layer of the city. Buildings change use and ownership over decades, but the configuration of streets endures for centuries, and it is on this configuration that urban life is organised. The space-syntax tradition has demonstrated, across a large body of empirical work, that the *topological* properties of the street network - not merely its geometry - predict pedestrian and vehicular movement, the location of retail and services, and patterns of social co-presence [1, 2]. Jane Jacobs' earlier, qualitative account of the conditions for urban vitality - a fine grain, mixed primary uses, short blocks and a dense web of accessible streets - anticipated much of this in the language of "organised complexity" [3].

The analytical toolkit that space syntax brought to this problem reduces a street's situation to a small set of scalar centralities. *Integration* measures topological closeness - how few turns separate a street from all others. *Choice* measures betweenness - how often a street lies on the shortest path between other streets. *Connectivity* counts a street's direct topological neighbours. These measures are powerful and have been validated repeatedly, but they are atomistic: each collapses the whole structure of a street's local network context into a single number describing one aspect of its centrality. They tell us *how central* a street is; they do not tell us *what kind of morphological context it sits within*.

This distinction matters because morphological context is precisely what varies across the urban fabric while centrality can hold constant. Two strokes may share an identical integration score yet belong to entirely different tissue: one embedded in a fine-grained nineteenth-century perimeter-block grid, the other a spine in a post-war dendritic housing estate. The classical metrics cannot separate these conditions, because the information that distinguishes them is not in any single centrality value but in the *shape* of the local connectivity structure - the way paths branch, recombine and close into loops in the immediate neighbourhood.

### 1.1 The gap: encoding morphological shape

What is missing is a descriptor that captures this shape directly. To be useful within and beyond space syntax, such a descriptor should satisfy four requirements. It should (i) operate on the dual graph of the street network, preserving theoretical continuity with space syntax's treatment of streets-as-nodes; (ii) encode the full structure of a node's local neighbourhood rather than a single centrality coefficient; (iii) be fixed-dimensional and comparable across locations and cities without bespoke calibration; and (iv) be directly consumable by modern machine-learning architectures, which increasingly operate on graphs.

The eigenvalue spectrum of a graph's Laplacian is a natural candidate for exactly this purpose. The Laplacian spectrum is a classical shape descriptor in spectral graph theory: it is invariant to node relabelling, can be normalised to a common scale, and compactly summarises connectivity structure - the number of components, the presence of bottlenecks, the degree of regularity. Crucially for the venue of this paper, the spectrum admits an information-theoretic reading: the *entropy* of the eigenvalue distribution measures how evenly structural "energy" is distributed across scales, and thus serves as a single-number proxy for morphological complexity.

### 1.2 Related work

Spectral descriptors of graphs are well established outside urban science. Laplacian and adjacency spectra are used as molecular fingerprints in chemistry and as parcellation signatures in brain connectomics, and the von Neumann (spectral) entropy of a graph, derived from the normalised Laplacian spectrum, is a standard measure of structural complexity in network science [4]. The second-smallest Laplacian eigenvalue - the Fiedler value, or algebraic connectivity - is a textbook measure of how close a graph is to disconnection [5]. These tools have not, however, been applied as *local, per-node* descriptors on the street dual graph.

A parallel line of work develops graph-level spectral descriptors as compact, comparable signatures. The heat- and wave-kernel signatures encode the Laplacian spectrum through diffusion and quantum-mechanical operators respectively [6, 7], and NetLSD derives a permutation- and size-invariant graph signature from the heat trace of the normalised Laplacian [8]. These descriptors are deliberately invariant to graph size by construction, so that graphs of different orders can be compared directly. Our descriptor differs in two respects. First, it is local and per-node: a fingerprint is computed on each stroke's k-hop ego subgraph rather than on a whole graph, which makes it a spatial field over the street network rather than a single whole-city signature. Second, rather than removing the size dependence analytically, we retain the raw spectral entropy and strip the ego-subgraph-size confound by regression (Section 2.4) - an approach that both isolates the residual morphological signal and quantifies, rather than assumes away, how much of the raw statistic that confound explains.

Entropy has been used in street-network analysis, but in different forms. Boeing computes the entropy of the *orientation* distribution of street segments as a measure of a city's grid-like regularity [9]; Mohajeri and Gudmundsson relate the entropies of street networks to their scaling and ordering [10, 11], and, more recently, Coutrot and colleagues link street-network entropy to human spatial behaviour [12]. These are geometric or city-aggregate constructs. Our contribution differs on both counts: it is *spectral* rather than orientation-based, and it is a *per-node neighbourhood descriptor* rather than a single value per city. Separately, the urban-morphometrics literature has produced rich geometric descriptors of plots and tessellations [13], and the network analysis of streets via the dual/primal graph is well developed [14], but neither encodes the local Laplacian spectrum.

### 1.3 Contribution

This paper makes four contributions.

1. We introduce the **size-adjusted spectral fingerprint**: a fixed-dimensional kernel-density encoding of the local graph-Laplacian spectrum on the COINS dual street graph. This is the primary, novel object.
2. Two interpretable scalar readouts are derived - the **Mesh Index** (normalised spectral entropy) and the **Connectivity Resilience Index** (Fiedler value) - and show that a size-adjustment step is *necessary*, not cosmetic: without it, the metrics are dominated by ego-subgraph size and collapse into disguised centrality proxies.
3. We characterise the descriptor empirically on a complete city network (Poznań), reporting its spatial structure, its near-orthogonality to classical integration, its capacity to separate morphological tissue types without supervision, and a comprehensive analysis of robustness to the method's three hyperparameters.
4. Finally, the method's scope is honestly delineated by demonstrating the descriptor is associated with neighbourhood-scale functional *diversity* but does not predict apartment price at street resolution - a clean negative result that establishes where the method does and does not apply.

The remainder of this paper is outlined as follows. Section 2 details the graph construction and descriptor methodology. Following this, Section 3 maps the results onto the Poznań network, while Section 4 unpacks the theoretical implications before we conclude.

---

## 2. Materials and Methods

### 2.1 Study area and graph construction

The study area is the city of Poznań, Poland, captured within a 12 × 12 km bounding box centred at 52.4082° N, 16.9336° E. Poznań is well suited to a morphological study because it concentrates several distinct fabric types within a single contiguous network: a medieval core (Stare Miasto and Ostrów Tumski), regular nineteenth-century perimeter-block grids (Łazarz, Jeżyce, Wilda), large socialist-era housing estates with dendritic and superblock layouts (Rataje, Piątkowo, Winogrady), and low-density post-1989 suburban development on the fringe.

Street centrelines were obtained from OpenStreetMap and reduced to a walkable network. Fragmented OSM segments were merged into perceptually continuous *strokes* using the COINS (Continuity in Street Networks) algorithm with an angular threshold of 30° [15]. We then construct the **dual graph** $G = (V, E)$: each stroke becomes a node, and an edge connects two nodes whenever the corresponding strokes intersect. This dual representation converts the geometric street map into a purely topological connectivity problem and maintains continuity with the space-syntax treatment of streets as the primitive units of analysis.

The resulting network comprises **1,908 stroke-nodes and 4,340 edges**, with a mean nodal degree of 4.55, spanning 42 administrative districts. Summary statistics are given in Table 1, and the network as a whole is shown in Figure 1.

**Table 1. Network and data summary (Poznań, 12 km).**

| Property | Value |
|---|---|
| Stroke-nodes (dual graph) | 1,908 |
| Edges | 4,340 |
| Mean degree (connectivity) | 4.55 (SD 6.06; median 3, max 99) |
| Graph density | 0.0024 |
| Connected components | 1 (fully connected) |
| Network diameter (topological steps) | 15 |
| Administrative districts | 42 |
| Integration (global), mean | 0.159 (SD 0.023) |
| Integration (local, R=3), mean | 0.375 (SD 0.022) |
| Choice (betweenness), mean | 0.0028 (SD 0.017) |
| Strokes with POI data (OSM primary / Google compar.) | 807 / 1,908 (42%) \| Google 1,739 (91%) |
| Strokes with ≥2 POIs (OSM / Google) | 555 / 1,440 |
| RCN property transactions (2016–2026) | 88,510 |

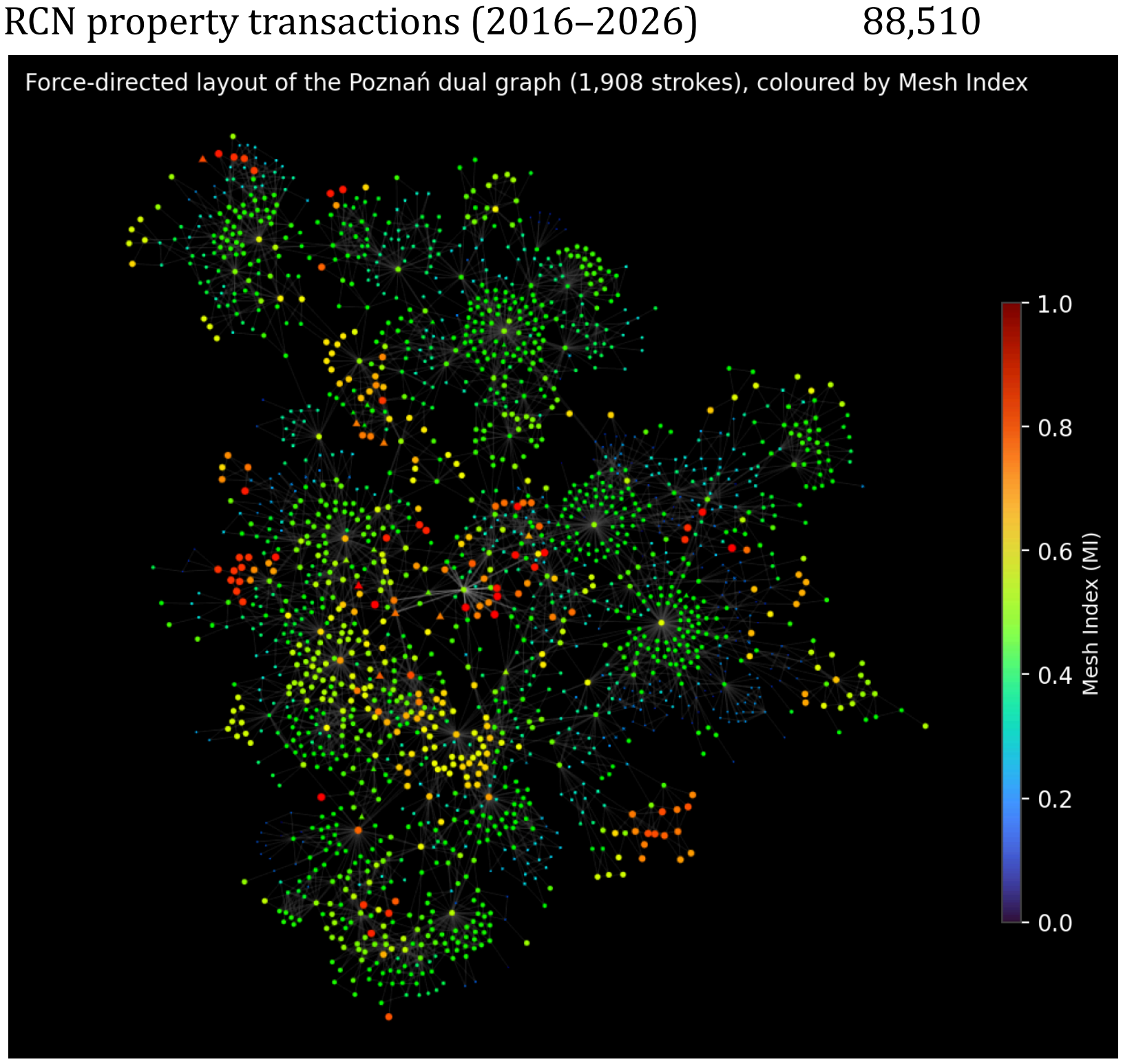

**Figure 1.** Force-directed (spring) layout of the Poznań dual graph - 1,908 stroke-nodes, with edges where strokes intersect - nodes coloured by Mesh Index and sized by connectivity. The organic, multi-cluster structure reflects the city's morphological heterogeneity; this is the interactive view used to explore the dual graph during the analysis.

*Alt text: A force-directed (spring) layout of the Poznań street dual graph: 1,908 stroke-nodes joined by edges where strokes intersect, drawn as an organic, multi-cluster cloud. Node colour encodes the Mesh Index and node size encodes connectivity, so denser, higher-connectivity regions appear as larger, distinctly coloured clusters.*

### 2.2 Classical space-syntax baselines

For comparison we compute four standard configurational metrics on the dual graph, following sDNA conventions [16]: global integration (network closeness centrality), local integration at a topological radius of three steps, choice (betweenness centrality), and connectivity (nodal degree). These constitute the baseline against which the spectral descriptor is assessed; they are not part of the contribution.

### 2.3 The spectral fingerprint

The spectral fingerprint encodes the local connectivity structure around each node as the shape of its Laplacian eigenvalue distribution. The procedure is given as Algorithm 1.

```
Algorithm 1 - Spectral fingerprint of node i
Input: dual graph G; node i; ego radius k; bins D; KDE bandwidth h
1. G_i ← k-hop ego subgraph of i in G (if |G_i| < 4, return null
fingerprint)
2. L ← D_deg − A                              (combinatorial Laplacian of
G_i)
3. {λ_1 ≤ λ_2 ≤ … ≤ λ_n} ← eigenvalues(L)
4. λ̂_j ← λ_j / λ_max                             (normalise spectrum to
[0,1])
5. fit Gaussian KDE with bandwidth h over {λ̂_j}
6. sample the KDE at D evenly spaced points in [0,1] → fingerprint φ_i ∈
ℝ^D
```

For each node $i$, we extract its $k$-hop ego subgraph $G_i$ - all nodes within $k$ topological steps, with the induced edges - and compute the combinatorial Laplacian $L = D_{deg} - A$, where $D_{deg}$ is the degree matrix and $A$ the adjacency matrix. The eigenvalues of $L$ are normalised by the largest eigenvalue to lie in [0, 1], which renders subgraphs of different sizes comparable. Because the eigenvalue set varies in length from node to node, we convert it to a fixed-dimensional vector by fitting a Gaussian kernel density estimate over the normalised eigenvalues and sampling it at $D$ evenly spaced points. The result, $\boldsymbol{\varphi_i} \in \mathbb{R}^{\mathbf{D}}$, is the spectral fingerprint of node $i$. Stacking all fingerprints yields an $N \times D$ matrix that is the structural layer of the descriptor.

The default parameters are $k = 3$, $D = 32$ and $h = 0.08$, with a minimum ego-subgraph size of four nodes below which a fingerprint is not computed. The choice of $k = 3$ is justified empirically in Section 3.5; the choices of $D$ and $h$ are shown there to be immaterial. Figure 2 illustrates the full pipeline from stroke to fingerprint to scalar readouts.

### 2.4 The Mesh Index and the size adjustment

This scalar readout summarises the fingerprint by its information content. The raw Mesh Index is the normalised Shannon entropy of the fingerprint,

$$MI_raw(i) = -\left( \Sigma_j\, p_j \log_2 p_j \right) / \log_2 D, \text{ with } p_j = \varphi_i[j] \,/\, \Sigma_j\, \varphi_i[j], \tag{1}$$

so that $MI_raw \in [0, 1]$. A high value indicates a distributed eigenvalue spectrum - a complex, multi-scale local connectivity structure - while a low value indicates a spectrum dominated by a few eigenvalues, characteristic of regular or monotonous topology.

Raw spectral entropy, however, is mechanically confounded by the size of the ego subgraph. Dense, well-connected cores generate large subgraphs whose normalised spectra concentrate, lowering entropy; sparse peripheral subgraphs generate flatter spectra with higher entropy. On the Poznań network this confound is severe: ego-subgraph size correlates with global integration at $r = 0.86$, and as a direct consequence raw MI correlates with integration at $r = -0.61$. Left uncorrected, the metric would be little more than an inverse restatement of centrality.

We therefore define the **Mesh Index** as the size-adjusted residual of MI_raw. We regress MI_raw on the logarithm of ego-subgraph size across all nodes, take the residual - the portion of a node's spectral entropy not explained by how large its neighbourhood is - and rescale the residuals to [0, 1]:

$$MI(i) = \mathrm{rescale}_{(0,1)}\left( MI_raw(i) - [a \cdot \log n_ego(i) + b] \right). \quad (2)$$

This adjustment is the methodological core of the metric. After it, the Mesh Index is near-orthogonal to integration ($r = 0.06$; Section 3.4), measuring local morphological complexity that is genuinely distinct from centrality rather than a disguised function of it.

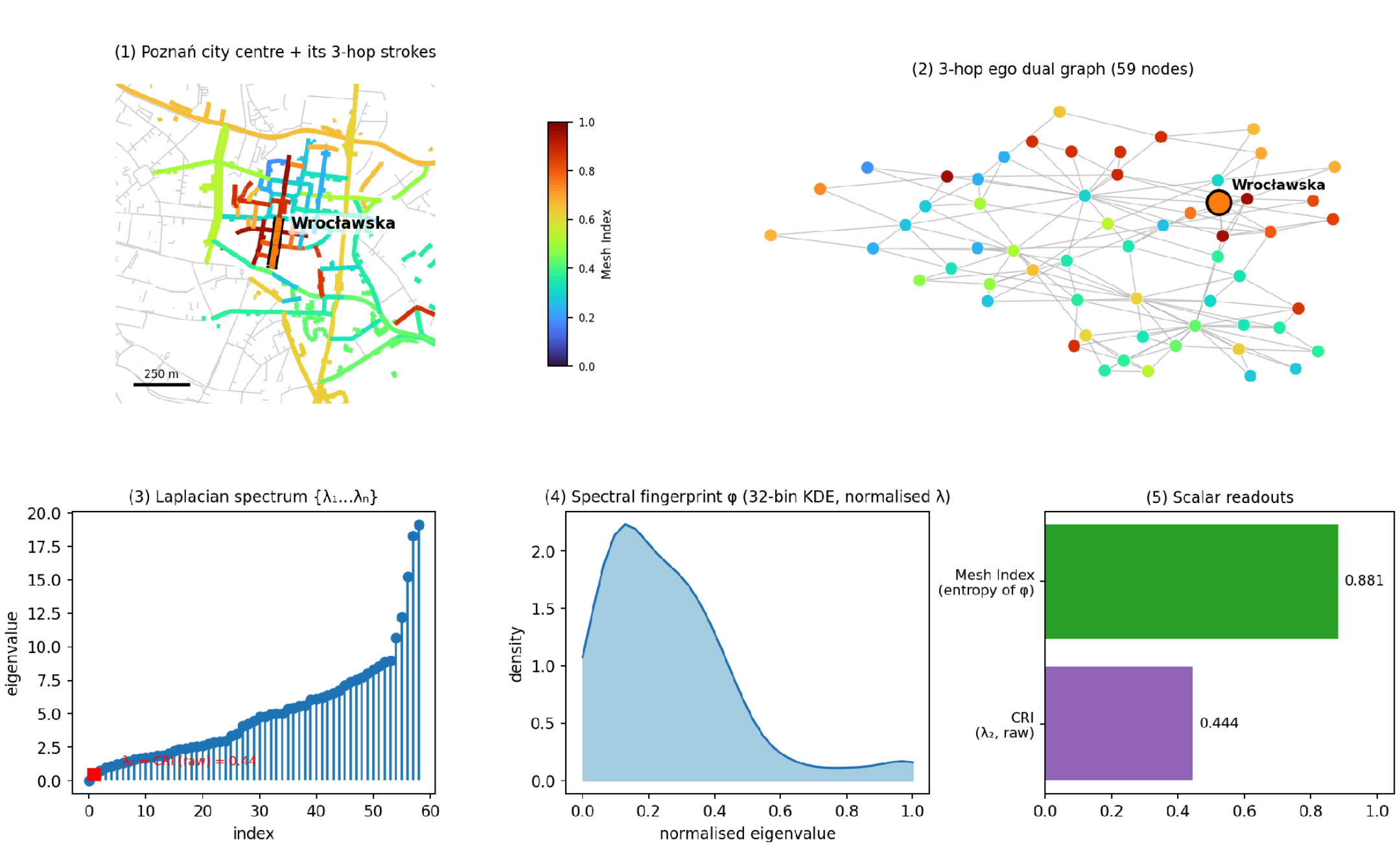

**Figure 2.** From a real street (Wrocławska, Poznań) to its spectral fingerprint and scalar readouts. (1) the 3-hop street neighbourhood rendered on the Poznań map, coloured by Mesh Index, with the focal stroke in orange; (2) the 3-hop ego dual graph (59 stroke-nodes; focal stroke in orange); (3) the eigenvalue spectrum of its combinatorial Laplacian, with the Fiedler value $\lambda_2$ (the basis of CRI) marked; (4) the 32-bin kernel-density fingerprint φ over the normalised spectrum; (5) the two scalar readouts derived from it - the Mesh Index (entropy of φ) and CRI.

*Alt text: A five-panel pipeline for one focal street (Wrocławska, Poznań). Panel 1 shows its 3-hop neighbourhood on the city map coloured by Mesh Index, with the focal stroke in orange; panel 2 the 3-hop ego dual graph of 59 stroke-nodes; panel 3 the combinatorial-Laplacian eigenvalue spectrum with the Fiedler value $\lambda_2$ marked; panel 4 the 32-bin kernel-density fingerprint over the normalised spectrum; panel 5 the two derived scalar readouts, the Mesh Index and CRI.*

Computationally, the per-node step is dominated by the dense eigendecomposition of the ego subgraph, which is $O(n_ego^3)$. Because the eigenvalues depend only on $k$, while $D$ and $h$ affect only the inexpensive kernel-density step, the spectrum can be computed once per radius and reused across all encoding choices. The full 12 km network is processed in seconds to a few minutes per radius on a single workstation, confirming city-scale feasibility.

### 2.5 The Connectivity Resilience Index

The second scalar captures local robustness rather than complexity. The Connectivity Resilience Index is the size-adjusted Fiedler value - the second-smallest Laplacian eigenvalue, $\lambda_2$, of the ego subgraph, which is the algebraic connectivity of the local network [5]:

$$\text{CRI_raw}(i) = \lambda_2(L_i) \tag{3}$$

$$\text{CRI}(i) = \text{rescale}_{(0,1)}(\ \text{CRI_raw}(i) - [a' \cdot \log n_ego(i) + b']\ ). \tag{4}$$

Algebraic connectivity is small when a graph is close to splitting into two weakly linked parts and large when it is robustly interconnected, a relationship made precise by Cheeger's inequality. As an urban descriptor, a low CRI marks a stroke whose local mesh depends on a bottleneck - a bridge or single corridor whose removal would fragment the neighbourhood. We stress the honest novelty boundary here: the Fiedler value is a textbook quantity; the contribution is its size-adjusted, per-node application to street morphology, not the operator itself. At very small radii the Fiedler value of tiny ego subgraphs can be degenerate, a caveat we return to in Section 4.

### 2.6 Validation data

Two external data layers are used to interpret the metrics; neither is an input to their computation. Point-of-interest (POI) data are drawn from two independent sources. The primary source is OpenStreetMap (OSM), an open, ODbL-licensed dataset queried via osmnx; POIs tagged amenity, shop, leisure or tourism are mapped to the eight standardised categories listed below through a fixed tag crosswalk (Appendix A.3) and assigned to the nearest stroke within 100 m, covering 807 of 1,908 strokes with at least one POI (555 with at least two). As a second, proprietary comparison we also query the Google Places API over the same eight categories - shops, restaurants, cafés, bars, pharmacies, culture, hotels and services - using an adaptive per-stroke radius capped at twenty results. Being closed and version-dependent, the Google layer is not reproducible and is used only to test whether the result is robust to the choice of data source; it covers 1,739 strokes (1,440 with at least two POIs). From each source we define a **functional-diversity** target: the Shannon entropy of the eight-category POI mix on each stroke, computed for the strokes carrying at least two POIs (OSM n = 555; Google n = 1,440). For the scope analysis only, we use residential transaction prices from the Polish property register (RCN, *Rejestr Cen Nieruchomości*), comprising 88,510 transactions for Poznań over 2016–2026, reduced to a median price per square metre per stroke. District boundaries from OpenStreetMap support the spatial cross-validation in Section 3.6.

## 3. Results

### 3.1 Spatial structure of the spectral metrics

The spectral metrics are spatially coherent and visibly different from classical centrality. Figure 3 maps the Mesh Index, the Connectivity Resilience Index, integration and connectivity across Poznań. The Mesh Index is highest in the dense nineteenth-century perimeter-block grids (the medieval core itself, with its large homogeneous mesh, registers low MI despite its centrality), and lowest in the dendritic peripheral estates, consistent with the interpretation of MI as a measure of local morphological complexity. The CRI map differs in pattern from integration: rather than tracing the globally central corridors, elevated CRI appears at well-meshed junctions and ring structures, reflecting local robustness rather than citywide closeness.

### 3.2 Complementarity to space syntax

Table 2 reports the correlations between the spectral readouts and the classical metrics. The size-adjusted Mesh Index is essentially uncorrelated with global integration (r = 0.06) and only modestly related to connectivity (r = 0.21), confirming that it carries information largely distinct from existing centrality measures. The one exception is local integration at radius three, with which MI shares moderate variance (r = 0.55) - unsurprising, since both are computed over the same three-step neighbourhood - though MI still contributes substantial independent information beyond it. We note explicitly that this near-orthogonality to global centrality is a property of the *size-adjusted* metric; the raw entropy is strongly (and spuriously) anti-correlated with integration, as Section 3.4 details.

**Table 2. Correlations of the spectral readouts with classical space-syntax metrics (size-adjusted; Pearson r, n = 1,908).**

| | Integration (global) | Integration (local, R=3) | Choice | Connectivity |
|---|---|---|---|---|
| Mesh Index (MI) | 0.06 | 0.55 | 0.14 | 0.21 |
| CRI | −0.02 | −0.13 | −0.02 | −0.03 |

The Mesh Index and CRI are themselves nearly uncorrelated (r = −0.01), confirming that they capture independent dimensions - **local complexity and local resilience** respectively. CRI is essentially uncorrelated with every classical metric. MI is near-orthogonal to *global* integration, choice and degree; its one moderate association is with *local* integration at radius three (r = 0.55), which is expected because both summarise the same three-step neighbourhood - yet MI still retains roughly 70% independent variance even relative to that measure.

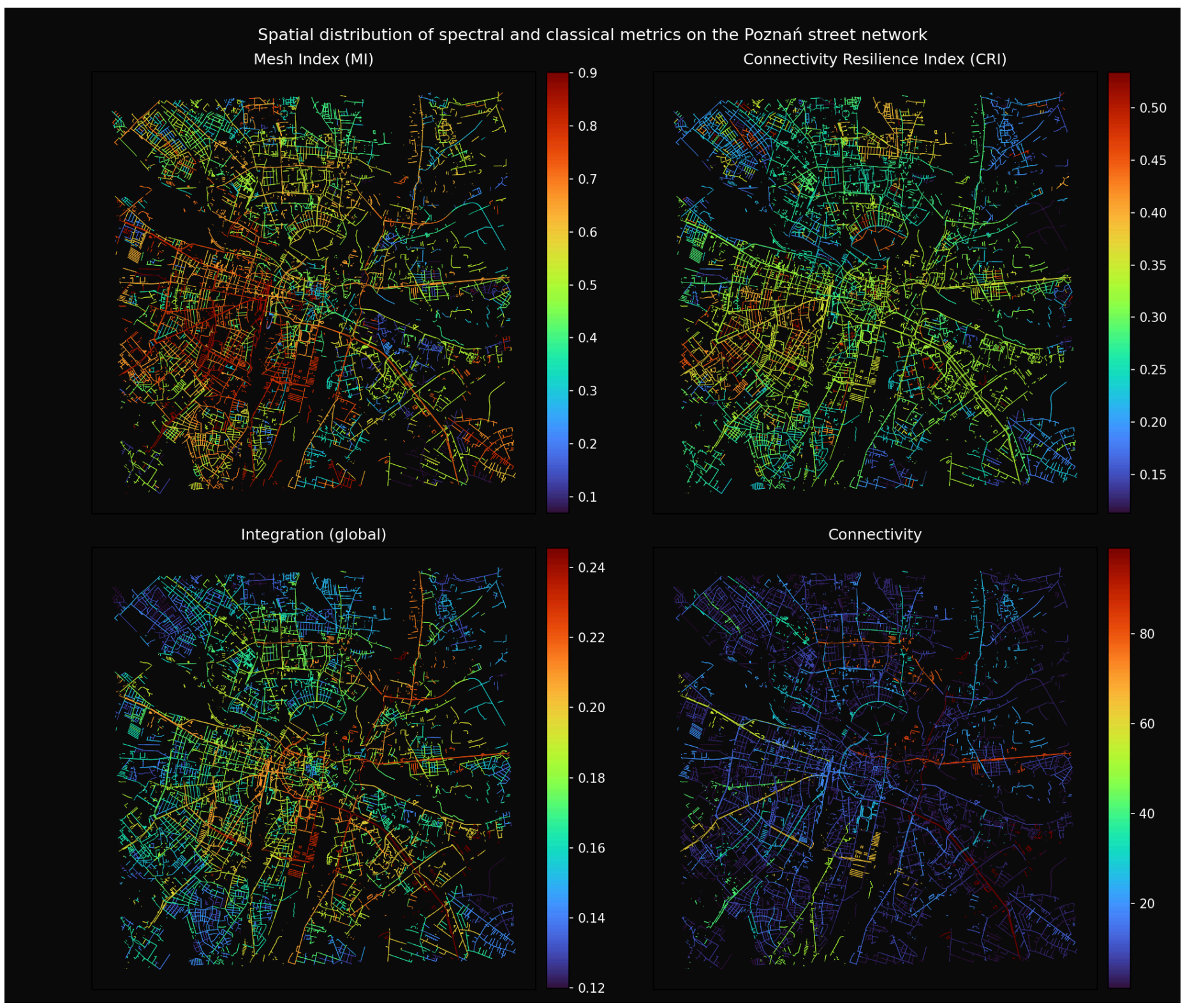

**Figure 3.** Spatial distribution of (a) Mesh Index, (b) Connectivity Resilience Index, (c) global integration and (d) connectivity rendered on the Poznań street geometry. Each line is a street segment coloured by its stroke's metric value - the same rendering used in the author's Space Syntax Studio.

*Alt text: Four maps of the Poznań street network drawn on real street geometry, each line a street segment coloured by its stroke's value of (a) the Mesh Index, (b) the Connectivity Resilience Index, (c) global integration and (d) connectivity. The Mesh Index and CRI maps show spatial patterns that differ visibly from the integration and connectivity maps.*

### 3.3 Morphological typologies from the fingerprint

Beyond its scalar summaries, the fingerprint itself separates morphological tissue types without supervision. Projecting the 1,908 fingerprint vectors to two dimensions and clustering them recovers groupings that align with known fabric (Figure 4): three archetypal fingerprint shapes recur across the city - **a flat, near-uniform spectrum characteristic of dense grid tissue; a left-skewed spectrum with mass concentrated at low eigenvalues, characteristic of dendritic suburban layouts; and a bimodal spectrum with a mid-range gap, characteristic of hub-and-arterial structures.** Strokes from the same district tend to cluster together in fingerprint space, indicating morphological coherence within neighbourhoods, and the fingerprint distinguishes tissue types - for example grid streets from radial spokes - that can share similar integration scores. A k-means partition of the fingerprints into three groups is well separated (silhouette 0.62, Davies–Bouldin 0.77), with sizes of roughly 1,500 (dendritic/suburban), 330 (grid/dense-urban) and 80 (hub/arterial) strokes.

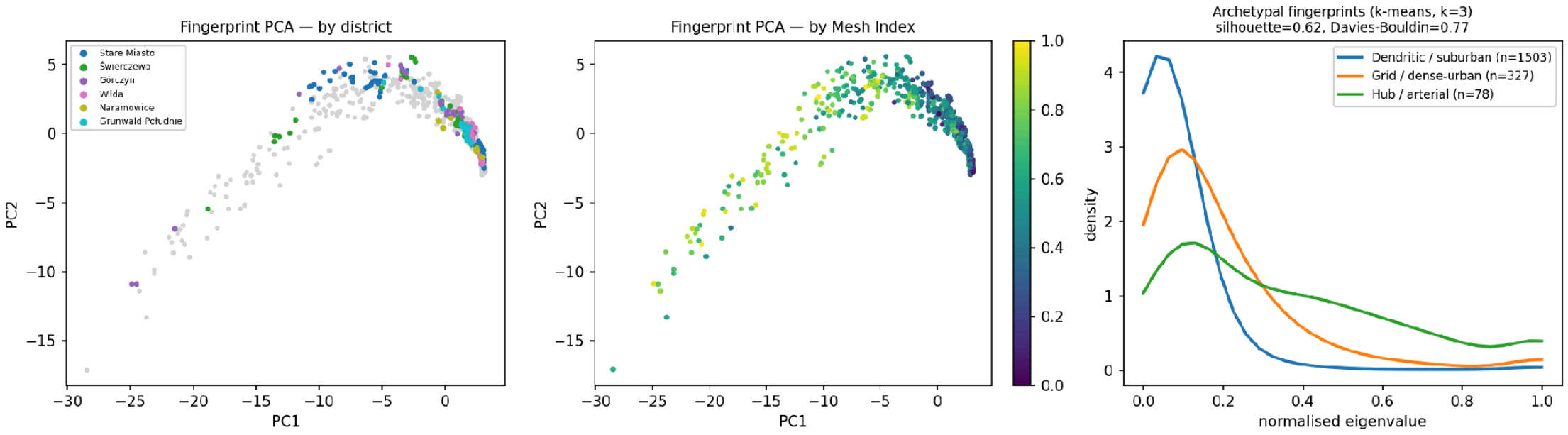

**Figure 4.** (a) Principal-component embedding of the 1,908 spectral fingerprints coloured by district; (b) the same embedding coloured by Mesh Index, showing a smooth gradient along the principal axis; (c) the three archetypal fingerprint shapes recovered by k-means (k = 3; silhouette 0.62): a dendritic/suburban type with mass concentrated at low eigenvalues, a grid/dense-urban type, and a flatter hub/arterial type.

*Alt text: Three panels. Panels (a) and (b) show a principal-component scatter embedding of the 1,908 spectral fingerprints, coloured by district in (a) and by Mesh Index in (b), where (b) reveals a smooth gradient along the principal axis. Panel (c) plots the three archetypal fingerprint curves from k-means (k = 3): a dendritic/suburban type with mass at low eigenvalues, a grid/dense-urban type, and a flatter hub/arterial type.*

### 3.4 The size adjustment is necessary

The decisive methodological result concerns the size adjustment introduced in Section 2.4. Figure 5 makes the case in three panels. Ego-subgraph size is strongly correlated with global integration (r = 0.86): central streets simply have larger local neighbourhoods. Because raw spectral entropy falls as subgraphs grow, the *raw* Mesh Index inherits this dependency and correlates with integration at r = −0.61 - it is, before correction, a disguised inverse-centrality measure. After regressing out the logarithm of ego-subgraph size, the residual Mesh Index is near-orthogonal to integration (r = 0.06). **The adjustment thus isolates the component of local spectral complexity that centrality does not already capture, and is the reason the Mesh Index constitutes a distinct quantity** rather than a re-expression of integration.

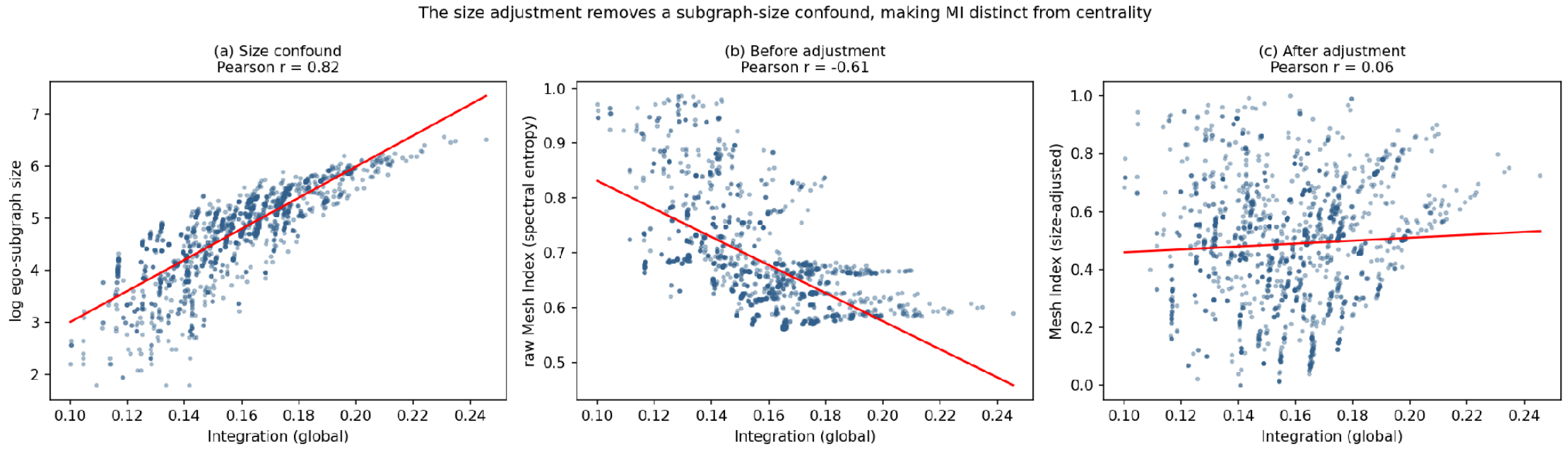

**Figure 5.** The necessity of the size adjustment. (a) Ego-subgraph size vs global integration (r = 0.86); (b) raw Mesh Index, i.e. spectral entropy, vs integration (r = −0.61); (c) size-adjusted Mesh Index vs integration (r = 0.06). Removing the subgraph-size confound leaves MI distinct from centrality rather than an inverse restatement of it.

*Alt text: Three scatter plots assessing the size adjustment. (a) ego-subgraph size against global integration, strongly positively correlated (r = 0.86); (b) the raw Mesh Index (spectral entropy) against integration, strongly*

*negative (r = −0.61); (c) the size-adjusted Mesh Index against integration, near-zero (r = 0.06). The adjustment removes the subgraph-size confound, leaving the Mesh Index independent of centrality.*

### 3.5 Robustness and scale: the k / D / bandwidth analysis

The descriptor has three hyperparameters: the ego radius $k$, the spectral resolution $D$ and the kernel bandwidth $h$. We computed the Mesh Index for the full Cartesian grid $k \in \{2, 3, 4, 5\} \times D \in \{16, 32, 64\} \times h \in \{0.05, 0.08, 0.12\}$ (36 settings) on the complete Poznań network, and compared node-level rankings using Spearman correlation. The full grid is reported in Appendix Table A1 and summarised in Figures 6 and 7. (As an internal check, the reimplementation used here reproduces the deployed metric exactly at the operating point, Spearman ρ = 1.0000.)

Two clear findings emerge. First, **spectral resolution and bandwidth are nuisance parameters**: at a fixed radius, varying $D$ across {16, 32, 64} and $h$ across {0.05, 0.08, 0.12} barely changes node rankings, with a minimum pairwise Spearman correlation of 0.961 at $k$ = 2 and 0.979 at the operating radius $k$ = 3, rising to 0.997 at $k$ = 5. The choropleths, clusters and correlations reported above are therefore not artefacts of the binning or smoothing choice. Figure 7 shows this directly: each row of the grid is flat across the $D$–$h$ columns.

Second, **the ego radius is a substantive scale parameter, not a nuisance**. Across radii the same node receives a materially different Mesh Index: relative to the $k$ = 3 metric, the Spearman correlation is −0.04 at $k$ = 2, 0.33 at $k$ = 4 and 0.18 at $k$ = 5. The metric computed at $k$ = 2 is, in effect, a different quantity, so the radius cannot be treated as an arbitrary default. This scale dependence is interpretable through the functional-diversity association (Section 3.6): the correlation between the Mesh Index and POI diversity is 0.17 at $k$ = 2 and 0.18 at $k$ = 3, but falls to 0.03 at $k$ = 4 and reverses to −0.05 at $k$ = 5. As the ego subgraph grows - mean size rising from 45 nodes at $k$ = 2 to 651 at $k$ = 5 - local morphology is diluted into citywide structure and the functional signal disappears. The neighbourhood scale at which morphology and use co-vary is therefore empirically bounded, and $k$ = 3 is both theoretically motivated and the radius at which the descriptor is most informative.

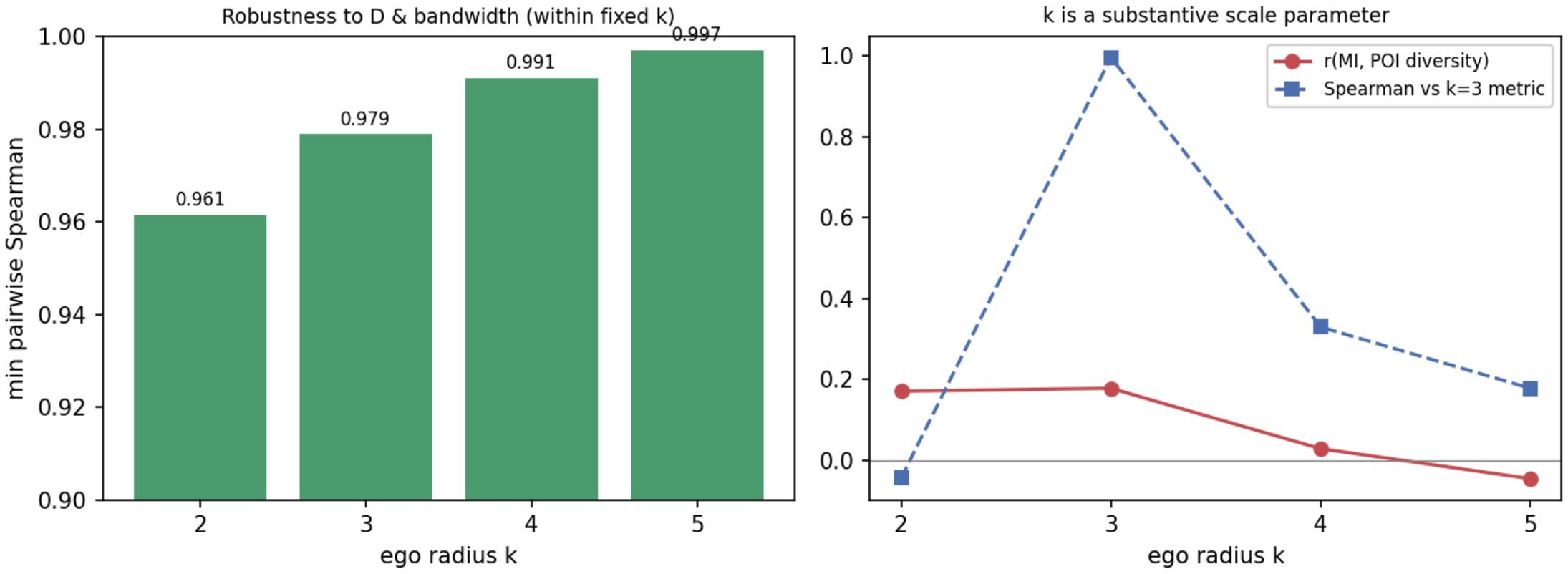

**Figure 6.** The ego radius is substantive while $D$ and bandwidth are not. Left: minimum pairwise Spearman correlation of the Mesh Index across the nine $D$–bandwidth combinations, by radius (all ≥ 0.96). Right: the functional-diversity correlation and the rank stability relative to $k$ = 3, as functions of radius.

*Alt text: Two line plots against ego radius. Left: the minimum pairwise Spearman correlation of the Mesh Index across the nine resolution–bandwidth combinations, all at or above 0.96. Right: the functional-diversity correlation and the rank stability relative to k = 3, each shown as a function of radius.*

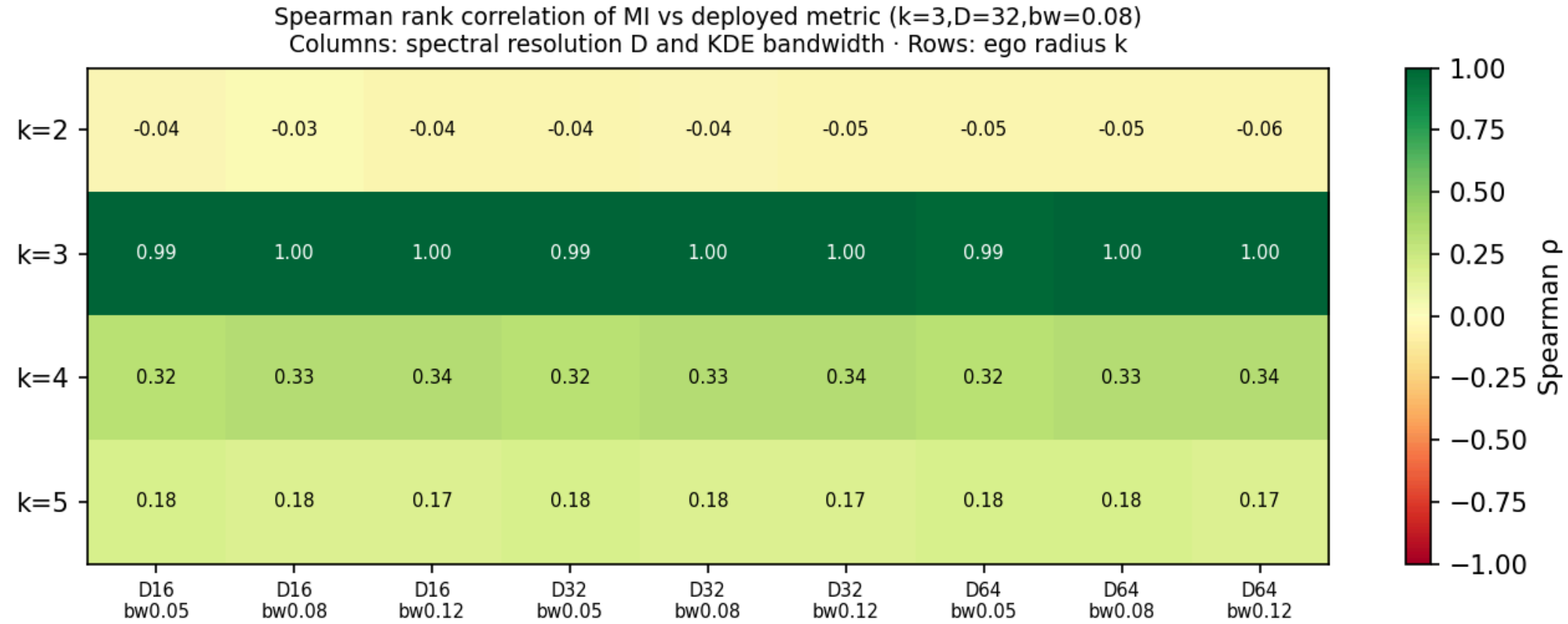

**Figure 7.** Spearman correlation of the Mesh Index against the deployed metric ($k$ = 3, $D$ = 32, $h$ = 0.08) across the full 4 × 9 grid. Rows (radius) differ substantially; columns (spectral resolution and bandwidth) are flat.

*Alt text: A 4 × 9 heatmap of Spearman correlations between the Mesh Index at each grid setting and the deployed metric (k = 3, D = 32, h = 0.08). Rows index ego radius and columns index spectral resolution and bandwidth; correlations vary substantially down the rows but stay flat across the columns.*

### 3.6 Functional diversity, honestly reported

We tested whether the structural descriptor relates to the *diversity* of street-adjacent activity - a relationship predicted by the natural-movement thesis, in which well-connected streets attract a varied mix of uses [2, 3]. The target is the Shannon entropy of the eight-category POI mix on each stroke with at least two POIs (OSM n = 555; Google n = 1,440), and the predictors are structural only (feature definitions and symbols are given in Table A2): no POI feature enters the model. Results are summarised in Table 3.

We measure functional diversity from two independent POI sources: OpenStreetMap (open and reproducible; our primary source) and the Google Places API (proprietary; a robustness comparison). On the open OSM data, functional diversity is positively and significantly associated with connectivity (r = 0.24), the Mesh Index (r = 0.19), integration (r = 0.18) and choice (r = 0.16) - reproducing, on fully open data, the associations on which this analysis rests. The proprietary Google layer behaves differently: its diversity is essentially unrelated to the Mesh Index (r = 0.05), connectivity (r = 0.02) and choice (r = 0.00), and instead tracks global integration (r = 0.21). The two sources agree only moderately on which streets are diverse (Pearson r = 0.25, Spearman ρ = 0.26 over the 491 strokes both cover). This divergence is interpretable, and informative: Google's rated, address-filtered commercial listings concentrate on globally central high streets, so they track centrality rather than the local morphological mix, whereas OSM's broader amenity coverage captures the everyday functional variety that the Mesh Index reflects. The morphology–diversity association is therefore a property of the open data and is, if anything, diluted by the proprietary source rather than created by it. In a multivariate model the same structural features explain a modest share of variance under a random train–test split ($R^2 \approx 0.11$, random forest), but this signal does not survive spatial cross-validation (below); we therefore do not attribute it to any single feature family.

Critically, the association is **within-district**. Under a leave-one-area-out spatial cross-validation (spatial blocks), the cross-sectional $R^2$ for the OSM target falls to approximately −0.06: a model trained on some parts of the city does not predict diversity in unseen ones. Two factors drive this - areas differ in baseline diversity according to their overall character, and the POI-covered subset is small once split spatially. The honest interpretation is that the Mesh Index ranks streets by their diversity-affording potential *relative to others in the same neighbourhood context*; it is a descriptor of relative position within a morphological setting, not an absolute predictor that transfers across the city.

**Table 3. Association between structural descriptors and POI functional diversity, by data source (OpenStreetMap - open, primary, n = 555; Google Places - proprietary, comparison, n = 1,440). Values are univariate Pearson r and held-out $R^2$.**

| Association with POI diversity | OSM (open, primary) | Google Places (proprietary) | Interpretation |
|---|---|---|---|
| Mesh Index (univariate r) | +0.19 | +0.05 | significant on OSM ($p<0.0001$) |
| Connectivity / choice / integration (r) | +0.24 / +0.16 / +0.18 | +0.02 / +0.00 / +0.21 | Google tracks integration only |
| Multivariate $R^2$ (sDNA+spectral, random) | **≈ 0.11** | **≈ 0.04** | **random-forest; modest** |
| Spatial-CV $R^2$ (leave-one-area-out) | −0.06 | +0.02 | no transfer across areas |

### 3.7 Scope delineation: what the metric does not predict

A clean negative result defines the method's boundary. We tested whether the spectral descriptor predicts residential property price - the median price per square metre per stroke, from 88,510 RCN transactions. Across every structural feature configuration the cross-sectional $R^2$ is at or below zero (Table 4), and augmenting the model with building-level context from the Polish topographic database (BDOT10k) - building function, floor counts, green-space and water proximity - does not change this. The reason is one of scale and resolution: apartment prices are driven principally by unit-level attributes (floor, condition, view, building age) that are invisible to any street-level spatial dataset, and the median across many buildings on a stroke averages away the variance these would explain. The descriptor characterises *what kind of activity a street's position affords*; it does not, and is not intended to, predict *the price that activity commands*. These are distinct questions at distinct spatial resolutions, and we regard stating the boundary clearly as part of the contribution.

**Table 4. Scope delineation - prediction of residential price (log median €/m² per stroke).**

| Feature set | $R^2$ (random split) |
|---|---|
| Spectral + sDNA | ≤ 0 |
| + BDOT10k building features | ≤ 0 |

# 4. Discussion

## 4.1 What the Mesh Index and CRI measure - and what they do not

The two readouts are purely topological summaries of a node's local network context. The Mesh Index is a proxy for the *complexity* of that context - the richness and multi-scale character of the routes available within a short topological reach - and, after the size adjustment, for the portion of that complexity not already captured by centrality. The CRI is a proxy for the *robustness* of the local mesh, distinguishing well-meshed neighbourhoods from those dependent on a bottleneck. Neither is a proxy for physical capacity, zoning, demographics, or - as Section 3.7 shows empirically - economic value. The size adjustment is what guarantees the first interpretation and rules out the second by construction.

## 4.2 Relation to space syntax

The spectral descriptor complements rather than competes with classical configurational analysis. Integration answers a global question - how reachable a street is from everywhere else - while the Mesh Index answers a local one - how spectrally complex the surrounding fabric is. The two can diverge sharply: a long radial arterial may be globally well integrated yet spectrally simple, while a fine-grained historic grid may be spectrally rich yet peripheral. Each condition implies a different functional expectation, and the near-orthogonality reported in Section 3.2 means the Mesh Index adds a dimension to the analyst's toolkit rather than duplicating an existing one.

## 4.3 An information-theoretic reading

Framing morphological complexity as the entropy of a local spectrum places the Mesh Index within the information-theoretic tradition and gives quantitative form to Jacobs' notion of organised complexity [3]. There is a conceptual parallel - which we offer as analogy rather than claim - to the diversity-and-ubiquity logic of economic complexity [17]: morphologically complex neighbourhoods, like economically complex economies, support a more varied repertoire of activity. The within-neighbourhood association between the Mesh Index and functional diversity is consistent with this reading.

## 4.4 Why the signal is local

The scale analysis of Section 3.5 gives an empirical answer to a question often left implicit in morphological studies: at what radius does the relationship between form and function hold? Here it is a neighbourhood-scale phenomenon, present at two to three topological steps and gone by four to five. This is consistent with how streets are used - people perceive and traverse their local area, and the grain and connective redundancy that afford a diverse mix of uses operate at that scale. For any cross-district application we recommend controlling for district fixed effects, in the same way hedonic models control for neighbourhood.

## 4.5 Limitations

Several limitations bound the present results. POI coverage is partial and uneven: the open OpenStreetMap layer used for the primary functional-diversity analysis covers 807 strokes with at least one POI (555 with at least two) and is best populated in central, well-mapped districts, so that analysis speaks to the covered subset rather than the whole city; the proprietary Google layer offers wider but non-reproducible coverage and is reported only as a robustness comparison. The study is single-city; while the descriptor is constructed to be comparable across cities, that comparability is not tested here. The radius $k$ remains a modelling decision, although Section 3.5 shows the other two hyperparameters do not matter and gives an empirical basis for $k = 3$. Finally, the COINS stroke segmentation is not stable across re-runs - connected-component ordering changes stroke identities - which currently obstructs longitudinal analysis on a fixed set of strokes and is the reason we make no temporal or causal claims in this paper. A stable identifier scheme anchored to street name and midpoint is the remedy and a prerequisite for future temporal work.

### 4.6 Future work

The descriptor is designed to extend. The immediate next steps are multi-city replication to test cross-city comparability, a stable-identifier scheme to enable longitudinal panels, and the integration of functional data layers and learned models to produce calibrated indices. These directions build on the structural foundation established here without altering the descriptive method itself.

---

## 5. Conclusions

We have introduced the spectral fingerprint - a fixed-dimensional, size-adjusted encoding of the local graph-Laplacian spectrum on the COINS dual street graph - together with two interpretable scalar readouts, the Mesh Index and the Connectivity Resilience Index. Applied to the complete street network of Poznań, the descriptor is robust to its encoding hyperparameters yet genuinely scale-dependent in its neighbourhood radius; its size adjustment is shown to be necessary, removing a strong subgraph-size confound and leaving the Mesh Index near-orthogonal to classical integration; the fingerprint separates morphological tissue types without supervision; and the Mesh Index is associated with the functional diversity of street-adjacent activity at the neighbourhood scale, while honestly not predicting property price. The result is a compact, robust and information-theoretic morphological descriptor that complements the space-syntax toolkit and is directly consumable by modern graph-learning pipelines, with an explicit and tested boundary on what it does and does not explain.

---

## Funding

This work received no external funding.

## Use of Generative AI and AI-Assisted Technologies

During the preparation of this work the author used a large language model (Claude, Anthropic) for two purposes: (i) language editing, reference formatting, and manuscript-structure review; and (ii) as a coding assistant to help implement the author's "Space Syntax Studio," an analytical tool designed by the author according to the author's own methodology and processing pipeline. All figures and data visualizations were produced by executing this author-developed software on the author's own data using reproducible computational methods (Python; osmnx, momepy, numpy, scipy, networkx). The author reviewed, verified, and edited all outputs, code, and text, and takes full responsibility for the content.

## Conflicts of Interest

I declare a competing interest: I am developing commercial urban-analytics software (under the working name 'Mesh Index Labs') and have filed a Polish patent application on a distinct, related land-screening system. These interests had no role in the study.

## Availability of Data and Materials

The data and code supporting the structural (spectral) findings of this study are openly available in Zenodo at https://doi.org/10.5281/zenodo.20968119 [18]. The deposit contains the sensitivity-analysis results (sensitivity_results.csv) and the spectral-fingerprint pipeline - ego-subgraph extraction, combinatorial-Laplacian spectrum, kernel-density encoding, and the Mesh Index and Connectivity Resilience Index computations - implemented in Python (numpy, scipy, networkx). Street-network geometry derives from OpenStreetMap (ODbL) and was processed with osmnx and momepy; the OpenStreetMap point-of-interest layer used for the primary functional-diversity analysis is likewise openly available. The Google Places point-of-interest layer is proprietary and subject to the Google Places API terms; it is not redistributed and is reported only as a robustness comparison. Residential transaction data are from the Polish Register of Real-Estate Prices (RCN, Rejestr Cen Nieruchomości), a public register. Software modules unrelated to the morphological descriptor reported here are not included in the release.

## Ethics

This study did not involve human participants, their identifiable data, or animals. It used only publicly available, aggregated datasets (OpenStreetMap; the Polish Register of Real-Estate Prices; the Google Places API). No ethical approval was required.

# Spectral Fingerprints of Street-Network Morphology: A Size-Adjusted Graph-Laplacian Descriptor of Urban Fabric

**Piotr C. Kamiński** *[Magdalena Abakanowicz University of the Arts: Poznań, PL; ORCID: 0009-0008-6573-2057] Correspondence: [piotr.kaminski@uap.edu.pl]*

---

## Appendix A. Supplementary material

**Table A1.** Full k / D / bandwidth sensitivity grid (36 settings): mean and standard deviation of the Mesh Index; Spearman correlation against the deployed metric; correlations of MI with integration and connectivity; correlation of MI with POI diversity; and ridge-regression $R^2$ for diversity. Provided as `sensitivity_results.csv`, available in Zenodo at https://doi.org/10.5281/zenodo.20968119.

**Table A2. Feature definitions.**

| Feature | Symbol | Definition | Range |
|---|---|---|---|
| Spectral fingerprint | $\varphi \in \mathbb{R}^{32}$ | 32-bin KDE of normalised Laplacian eigenvalues of the 3-hop ego subgraph | per-bin $\geq 0$ |
| Mesh Index | MI | size-adjusted normalised spectral entropy of $\varphi$ | [0, 1] |
| Connectivity Resilience Index | CRI | size-adjusted Fiedler value $\lambda_2$ of the ego Laplacian | [0, 1] |
| Integration (global) | I_g | network closeness centrality on the dual graph | (0, 1) |
| Integration (local, R=3) | I_3 | closeness within a 3-step topological radius | (0, 1) |
| Choice | C | betweenness centrality | $\geq 0$ |
| Connectivity | $\kappa$ | nodal degree in the dual graph | $\geq 1$ |

**Appendix A.3 - Reproducibility.** The structural pipeline relies on open data and open tools: OpenStreetMap centrelines (via osmnx), the COINS algorithm (momepy), and standard scientific Python (numpy, scipy, networkx) for the spectral computation. Point-of-interest data are subject to the Google Places API terms. Functional-diversity validation draws on two POI layers - OpenStreetMap (amenity, shop, leisure and tourism tags mapped to the eight categories through a fixed crosswalk; open and reproducible, used as the primary source) and the Google Places API (proprietary; reported only as a robustness comparison). A code-availability statement will specify which components are released. Spectral fingerprints, scalar readouts, and all derived visualizations were computed and rendered by the author's Space Syntax Studio, a Python application implementing the pipeline described in Section 2; its development was supported by an AI coding assistant.